# Electromagnetic binding and radiation force reversal on a pair of electrically conducting cylinders of arbitrary geometrical cross-section with smooth and corrugated surfaces

F.G. Mitri*

A B S T R A C T

The electromagnetic (EM) radiation force-per-length exerted on a pair of electrically-conducting cylindrical particles of circular and non-circular cross-sections is examined using a formal semi-analytical method based on boundary matching in cylindrical coordinates. Initially, the scattering coefficients of the particle pair are determined by imposing suitable boundary conditions leading linear systems of equations computed via matrix inversion and numerical integration procedures. Standard cylindrical (Bessel and Hankel) wave functions are used and closed-form expressions for the dimensionless longitudinal and transverse radiation force functions are evaluated assuming either magnetic (TE) or electric (TM) plane wave incidences. Particle pairs with smooth and corrugated surfaces are considered and numerical computations are performed with emphasis on the distance separating their centers of mass, the angle of incidence of the incident illuminating field and the surface roughness. Adequate convergence plots confirm the validity of the method to evaluate the radiation force functions, and the model is adaptable to any frequency range (i.e. Rayleigh, Mie or geometrical optics regimes). The results can find potential applications in optical tweezers and other related applications in fluid dynamics. In addition, the acoustical analogue is discussed.

## 1. Introduction

In a recent investigation [1] related to the electromagnetic (EM)/optical binding [2] between two circular electrically-conducting cylinders, a mathematical analytical model was developed for the determination of the scattering coefficients of the particles and used subsequently to compute the longitudinal and transverse radiation force components. The numerical predictions have anticipated the generation of an EM radiation force cancellation effect, as well as the emergence of repelling and attractive forces depending on the distance separating the particles, and the incident field parameters. The acoustic analog was also developed and similar effects have been analyzed [3]. The method was based on the partial-wave series expansion in cylindrical coordinates with appropriate boundary matching conditions that were enforced over each particle surface. In this technique, the mathematical basis is composed of cylindrical wave functions such as the standard Bessel and Hankel functions [4], which form a complete set suitable for describing accurately the scattering phenomenon with partial-wave series expansions.

When the cylinder geometrical cross-section deviates from the circular shape, the method has been proven useful to determine the internal and external EM fields for irregularly shaped particles in laser-particle interactions [5]. Moreover, other investigations related to the acoustical scattering, radiation force and torque on elliptical particles [6-10] were developed. Nonetheless, in certain cases, multiple particles are subjected to an incident illuminating field such that interparticle forces causing optical binding effects can arise, which do not exist in the single object case. Thus, it is of some importance to develop improved analytical modeling to predict the action of the EM forces that arise in due course [2, 11-16].

The aim of this work is therefore directed toward extending the scope of the previous analysis [1] for the case of a pair of irregularly shaped cylindrical particles from the standpoint of EM radiation force theory. This effect is intrinsically connected with the multiple EM scattering between the particles [17] with particular shapes that deviate from the circular geometry [18-22]. Here, the formalism based on the partial-wave series expansion method using standard cylindrical Bessel and Hankel wave functions [1] is utilized to examine the cases of a pair of cylindrical particle of irregular geometrical cross-section with or without corrugations. The formalism is applicable to any range of frequencies such that either the long- or short-wavelength with respect to the size of the particles can be examined rigorously. Notice also that apart from their practical importance, two-dimensional EM scattering and radiation force analytical models provide rigorous, convenient and efficient ways of displaying emergent physical phenomena without involving the mathematical/algebraic manipulations encountered in solving three-dimensional problems [23-29], and this work could assist in developing numerical codes for arbitrary shaped particles in 3D.

---

*Electronic mail: F.G.Mitri@ieee.org

## 2. Analysis of the multiple EM scattering for a pair of electrically-conducting arbitrary-shaped cylindrical particles in plane waves with TE and TM polarizations

Consider a pair of cylindrical particles with arbitrary geometrical cross-section (in 2D) where their centers of mass are separated by a distance $d$ as shown in Fig. 1. A particular type of such particles has a continuous surface shape function [30] expressed in the system of coordinates $(r, \theta, z)$ as,

$$A_{\theta,1} = \left[(\cos\theta/a_1)^2 + (\sin\theta/b_1)^2\right]^{-1/2} + d_1 \cos(\ell_1 \theta), \quad (1)$$

where $a_1$ and $b_1$ are characteristic semi-axes, $d_1$ is the amplitude of the surface roughness, and $\ell_1$ is an integer number which determines the periodicity of the surface corrugations.

Similarly, in the system of coordinates $(r', \theta', z')$, the surface shape function of the second particle is expressed as,

$$A_{\theta,2} = \left[(\cos\theta'/a_2)^2 + (\sin\theta'/b_2)^2\right]^{-1/2} + d_2 \cos(\ell_2 \theta'). \quad (2)$$

Assuming now an incident field of plane progressive waves on the pair of particles with a magnetic field polarized along the axial $z$-direction (known also as TE polarization where the axial component of the electric field vanishes), application of the boundary condition requires that the tangential component of the total (i.e., incident + scattered) electric field vector $\mathbf{E}_{\{1,2\}}$ vanishes [30] (at $r = A_{\theta,1}$, and $r' = A_{\theta,2}$) such that,

$$\left(\mathbf{n}_{\{1,2\}} \times \mathbf{E}_{\{1,2\}}\right) \cdot \mathbf{e}_{\{z,z'\}}\Big|_{\{r=A_{\theta,1}, r'=A_{\theta,2}\}} = 0, \quad (3)$$

where the normal vector is expressed as,

$$\mathbf{n}_1 = \mathbf{e}_r - \left(\frac{1}{A_{\theta,1}}\right)\frac{dA_{\theta,1}}{d\theta}\mathbf{e}_\theta,$$

$$\mathbf{n}_2 = \mathbf{e}_{r'} - \left(\frac{1}{A_{\theta,2}}\right)\frac{dA_{\theta,2}}{d\theta'}\mathbf{e}_{\theta'}, \quad (4)$$

with $\mathbf{e}_{\{r,r'\}}$ and $\mathbf{e}_{\{\theta,\theta'\}}$ denoting the outward unit vectors along the radial and tangential directions, respectively, and $\mathbf{e}_{\{z,z'\}}$ denoting the outward unit vector along the axial direction.

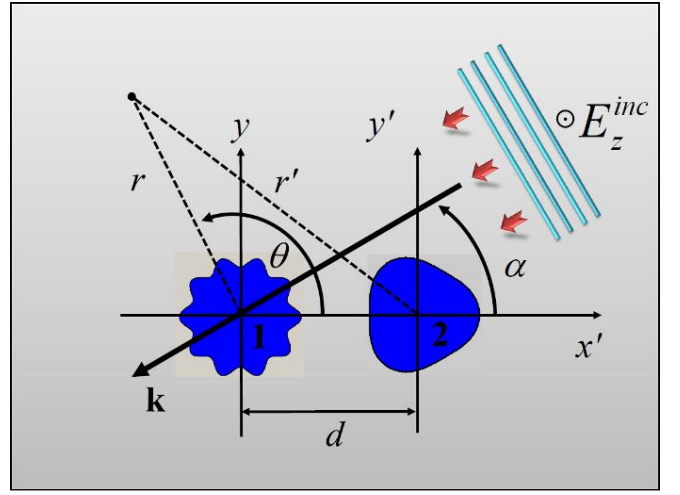

**Fig. 1.** A graphic describing the interaction of an incident electric field polarized along the axial $z$-direction perpendicular to the plane of the figure (known also as TM polarization) with a pair of arbitrary-shaped particles in 2D separated by a distance $d$. The cylindrical coordinate system $(r, \theta, z)$ is referenced to the center of the first particle while the primed one $(r', \theta', z')$ is located at the center of mass of the second one.

Following the theoretical development presented in the previous analysis [1], the expression for the *radial* component of the total electric field vector in the system of coordinates $(r, \theta, z)$ is expressed as,

$$E_{r,1}^{\text{tot}}(r,\theta,t)\Big|_{r<d} = E_0 e^{-i\omega t} \left[\sum_{n=-\infty}^{+\infty}\left(i\tfrac{n}{kr}\right)\binom{i^{-n}e^{-in\alpha}J_n(kr)}{+C_{n,1}H_n^{(1)}(kr)}e^{in\theta} \right.$$
$$\left. +\sum_{n=-\infty}^{+\infty}\left(i\tfrac{n}{kr'}\right)\left(C_{n,2}\sum_{m=-\infty}^{+\infty}J_m(kr)H_{m-n}^{(1)}(kd)e^{im\theta}\right)\right], \quad (5)$$

where $J_n(\cdot)$ and $H_n^{(1)}(\cdot)$ are the cylindrical Bessel and Hankel functions of the first kind, respectively, $k$ is the wavenumber and $d$ is the interparticle distance (Fig. 1).

Similarly, the polar component of the total electric field vector in the system of coordinates $(r, \theta, z)$ is expressed as,

$$E_{\theta,1}^{\text{tot}}(r,\theta,t)\Big|_{r<d} = -E_0 e^{-i\omega t} \left[\sum_{n=-\infty}^{+\infty}\binom{i^{-n}e^{-in\alpha}J_n'(kr)}{+C_{n,1}H_n^{(1)'}(kr)}e^{in\theta} \right.$$
$$\left. +\sum_{n=-\infty}^{+\infty}\left(C_{n,2}\sum_{m=-\infty}^{+\infty}J_m'(kr)H_{m-n}^{(1)}(kd)e^{im\theta}\right)\right]. \quad (6)$$

In the system of coordinates $(r', \theta', z')$, the radial and polar components of the total electric field vector are expressed, respectively, as,

$$E_{r,2}^{\text{tot}}(r',\theta',t)\Big|_{r'<d} = E_0 e^{-i\omega t}\left[\sum_{n=-\infty}^{+\infty}\left(i\frac{n}{kr'}\right)\begin{pmatrix}i^{-n}e^{-in\alpha}e^{-ikd\cos\alpha}J_n(kr')\\ +C_{n,2}H_n^{(1)}(kr')\end{pmatrix}e^{in\theta'}\right.$$
$$\left.+\sum_{n=-\infty}^{+\infty}\left(i\frac{n}{kr}\right)\left(C_{n,1}\sum_{m=-\infty}^{+\infty}J_m(kr')H_{n-m}^{(1)}(kd)e^{im\theta'}\right)\right],\quad(7)$$

$$E_{\theta,2}^{\text{tot}}(r',\theta',t)\Big|_{r'<d} = -E_0 e^{-i\omega t}\left[\sum_{n=-\infty}^{+\infty}\begin{pmatrix}i^{-n}e^{-in\alpha}e^{-ikd\cos\alpha}J_n'(kr')\\ +C_{n,2}H_n^{(1)'}(kr')\end{pmatrix}e^{in\theta'}\right.$$
$$\left.+\sum_{n=-\infty}^{+\infty}\left(C_{n,1}\sum_{m=-\infty}^{+\infty}J_m'(kr')H_{n-m}^{(1)}(kd)e^{im\theta'}\right)\right].\quad(8)$$

Application of the boundary condition given by Eq.(3) in both systems of coordinates leads to coupled systems of equations given as

$$\begin{cases}\sum_{\ell=-\infty}^{+\infty}\left[\psi_{\ell n}^{\text{TE}}+C_{n,1}^{\text{TE}}\Omega_{\ell n}^{\text{TE}}+C_{n,2}^{\text{TE}}\Omega_{\ell nm}^{\text{TE}}\right]=0,\\ \sum_{\ell'=-\infty}^{+\infty}\left[\psi_{\ell' n'}^{\text{TE}}+C_{n,1}^{\text{TE}}\Omega_{\ell' n'}^{\text{TE}}+C_{n,2}^{\text{TE}}\Omega_{\ell' n' m'}^{\text{TE}}\right]=0,\end{cases}\quad(9)$$

where,

$$\psi_{\ell n}^{\text{TE}}=\frac{1}{2\pi}\sum_{n=-\infty}^{+\infty}i^{-n}e^{-in\alpha}\int_0^{2\pi}\left[\begin{matrix}kJ_n'(kA_{\theta,1})\\ -i\left(\dfrac{n}{A_{\theta,1}^2}\right)\dfrac{dA_{\theta,1}}{d\theta}J_n(kA_{\theta,1})\end{matrix}\right]e^{i(n-\ell)\theta}d\theta,\quad(10)$$

$$\Omega_{\ell n}^{\text{TE}}=\frac{1}{2\pi}\sum_{n=-\infty}^{+\infty}\int_0^{2\pi}\left[\begin{matrix}kH_n^{(1)'}(kA_{\theta,1})\\ -i\left(\dfrac{n}{A_{\theta,1}^2}\right)\dfrac{dA_{\theta,1}}{d\theta}H_n^{(1)}(kA_{\theta,1})\end{matrix}\right]e^{i(n-\ell)\theta}d\theta,\quad(11)$$

$$\Omega_{\ell nm}^{\text{TE}}=\frac{1}{2\pi}\sum_{n=-\infty}^{+\infty}\sum_{m=-\infty}^{+\infty}H_{m-\ell}^{(1)}(kd)\int_0^{2\pi}\left[\begin{matrix}kJ_m'(kA_{\theta,1})\\ -i\left(\dfrac{\ell}{A_{\theta,1}\Delta_{\theta,1}}\right)\dfrac{dA_{\theta,1}}{d\theta}J_m(kA_{\theta,1})\end{matrix}\right]e^{i(m-n)\theta}d\theta,\quad(12)$$

$$\psi_{\ell' n'}^{\text{TE}}=\frac{1}{2\pi}\sum_{n'=-\infty}^{+\infty}i^{-n'}e^{-in'\alpha}e^{-ikd\cos\alpha}\int_0^{2\pi}\left[\begin{matrix}kJ_{n'}'(kA_{\theta,2})\\ -i\left(\dfrac{n'}{A_{\theta,2}^2}\right)\dfrac{dA_{\theta,2}}{d\theta'}J_{n'}(kA_{\theta,2})\end{matrix}\right]e^{i(n'-\ell')\theta'}d\theta',\quad(13)$$

$$\Omega_{\ell' n'}^{\text{TE}}=\frac{1}{2\pi}\sum_{n'=-\infty}^{+\infty}\int_0^{2\pi}\left[\begin{matrix}kH_{n'}^{(1)'}(kA_{\theta,2})\\ -i\left(\dfrac{n'}{A_{\theta,2}^2}\right)\dfrac{dA_{\theta,2}}{d\theta'}H_{n'}^{(1)}(kA_{\theta,2})\end{matrix}\right]e^{i(n'-\ell')\theta'}d\theta',\quad(14)$$

$$\Omega_{\ell' n' m'}^{\text{TE}}=\frac{1}{2\pi}\sum_{n'=-\infty}^{+\infty}\sum_{m'=-\infty}^{+\infty}H_{\ell'-m'}^{(1)}(kd)\int_0^{2\pi}\left[\begin{matrix}kJ_{m'}'(kA_{\theta,2})\\ -i\left(\dfrac{\ell'}{A_{\theta,2}\Delta_{\theta,2}}\right)\dfrac{dA_{\theta,2}}{d\theta'}J_{m'}(kA_{\theta,2})\end{matrix}\right]e^{i(m'-n')\theta'}d\theta',\quad(15)$$

where $\Delta_{\theta,1}=\sqrt{A_{\theta,1}^2+d^2-2A_{\theta,1}d\cos\theta}$, and $\Delta_{\theta,2}=\sqrt{A_{\theta,2}^2-d^2+2A_{\theta,2}d\cos\theta'}$.

Correspondingly, if an incident electric field polarized along the axial *z*-direction is considered (known as TM polarization), the boundary condition requiring that the tangential component of the total (i.e., incident + scattered) electric field vector $\mathbf{E}_{\{1,2\}}$ vanishes (at $r = A_{\theta,1}$, and $r' = A_{\theta,2}$) is expressed as,

$$\left(\mathbf{n}_{\{1,2\}}\times\mathbf{E}_{\{1,2\}}\right)\cdot\mathbf{e}_{\{\theta,\theta'\}}\Big|_{\{r=A_{\theta,1},r'=A_{\theta,2}\}}=0.\quad(16)$$

The expressions for the axial component of the total electric field vector in the system of coordinates $(r,\theta,z)$ and $(r',\theta',z')$ are expressed, respectively, as [1],

$$E_{z,1}^{\text{tot}}(r,\theta,t)\Big|_{r<d}=E_0 e^{-i\omega t}\left[\sum_{n=-\infty}^{+\infty}\begin{pmatrix}i^{-n}e^{-in\alpha}J_n(kr)\\ +C_{n,1}H_n^{(1)}(kr)\end{pmatrix}e^{in\theta}\right.$$
$$\left.+\sum_{n=-\infty}^{+\infty}\left(C_{n,2}\sum_{m=-\infty}^{+\infty}J_m(kr)H_{m-n}^{(1)}(kd)e^{im\theta}\right)\right],\quad(17)$$

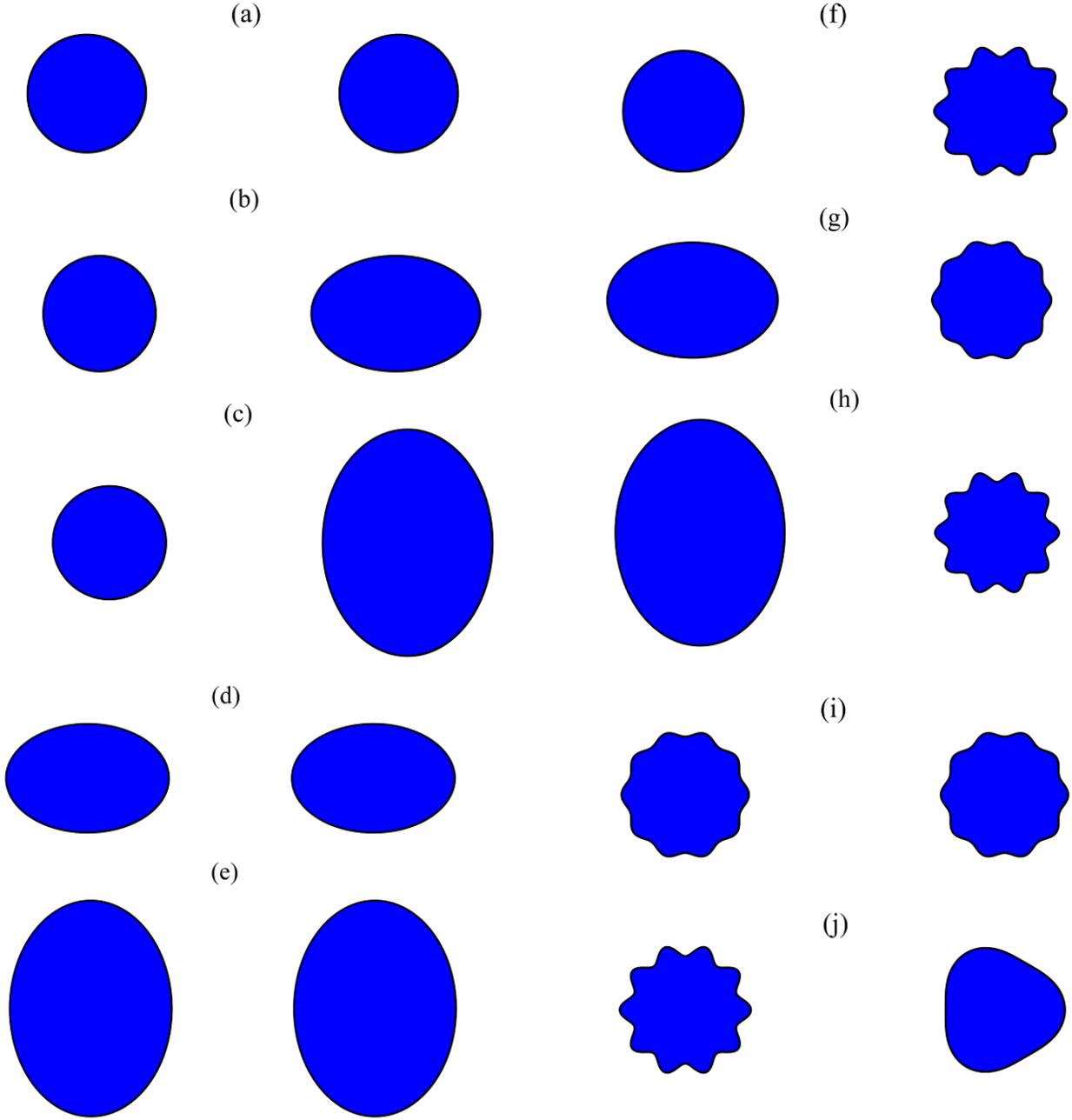

**Fig. 2.** Panels (a)-(e) display the geometrical cross-sections of the cylindrical particle pairs considered in this study with circular [i.e., panel (a)] and other non-circular forms with smooth boundaries. The particle in the left-hand side of each panel corresponds to object 1 in Fig. 1, while the one in the right-hand side is denoted as particle 2.

$$E_{z,2}^{tot}(r',\theta',t)\big|_{r'<d} = E_0 e^{-i\omega t}\left[\sum_{n=-\infty}^{+\infty}\begin{pmatrix}i^{-n}e^{-in\alpha}e^{-ikd\cos\alpha}J_n(kr')\\+C_{n,2}H_n^{(1)}(kr')\end{pmatrix}e^{in\theta'}\right.$$
$$\left.+\sum_{n=-\infty}^{+\infty}\left(C_{n,1}\sum_{m=-\infty}^{+\infty}J_m(kr')H_{n-m}^{(1)}(kd)e^{im\theta'}\right)\right].$$
(18)

**Fig. 2 (Continued).** Panels (f)-(j) display the geometrical cross-sections of the cylindrical particle pairs considered in this study with circular and non-circular forms with some objects having corrugated surfaces. The particle in the left-hand side of each panel corresponds to object 1 in Fig. 1, while the one in the right-hand side is denoted as particle 2.

Application of the boundary condition given by Eq.(16) in both systems of coordinates leads to coupled systems of equations given as

$$\begin{cases}\sum_{\ell=-\infty}^{+\infty}\left[\psi_{\ell n}^{TM}+C_{n,1}^{TM}\Omega_{\ell n}^{TM}+C_{n,2}^{TM}\Omega_{\ell nm}^{TM}\right]=0,\\ \sum_{\ell'=-\infty}^{+\infty}\left[\psi_{\ell'n'}^{TM}+C_{n,1}^{TM}\Omega_{\ell'n'}^{TM}+C_{n,2}^{TM}\Omega_{\ell'n'm'}^{TM}\right]=0,\end{cases}$$
(19)

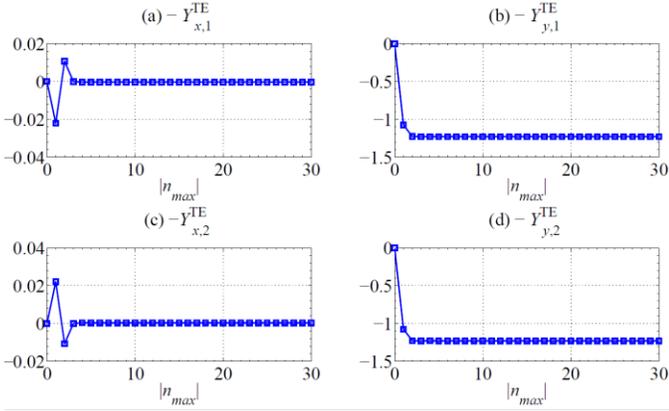

**Fig. 3.** Convergence plots for the longitudinal and transverse dimensionless radiation force functions assuming TE polarization versus the truncation order $|n_{max}|$ for a pair or circular cylinders with smooth surfaces as shown in panel (a) of Fig. 2.

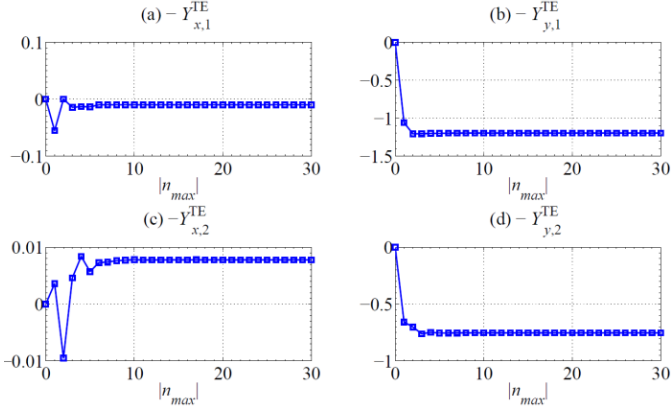

**Fig. 4.** The same as in Fig. 3, but for a circular and elliptical cylinders with smooth surfaces as shown in panel (b) of Fig. 2.

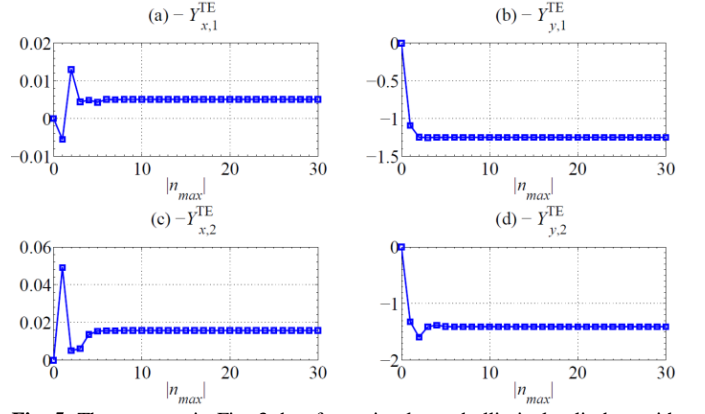

**Fig. 5.** The same as in Fig. 3, but for a circular and elliptical cylinders with smooth surfaces as shown in panel (c) of Fig. 2.

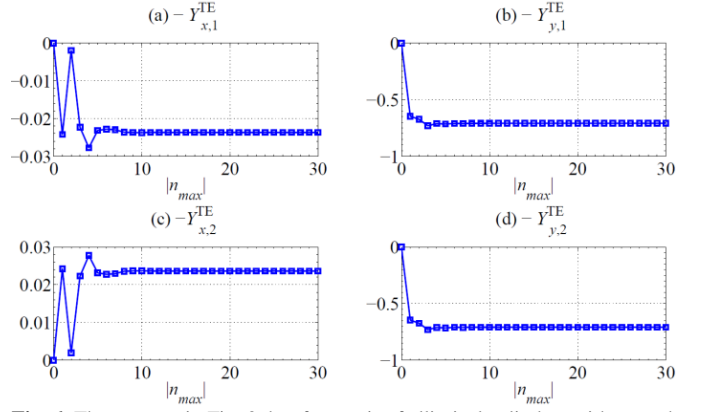

**Fig. 6.** The same as in Fig. 3, but for a pair of elliptical cylinders with smooth surfaces as shown in panel (d) of Fig. 2.

where,

$$\psi_{\ell n}^{TM} = \frac{1}{2\pi} \sum_{n=-\infty}^{+\infty} i^{-n} e^{-in\alpha} \int_0^{2\pi} J_n(kA_{\theta,1}) e^{i(n-\ell)\theta} d\theta, \quad (20)$$

$$\Omega_{\ell n}^{TM} = \frac{1}{2\pi} \sum_{n=-\infty}^{+\infty} \int_0^{2\pi} H_n^{(1)}(kA_{\theta,1}) e^{i(n-\ell)\theta} d\theta, \quad (21)$$

$$\Omega_{\ell n m}^{TM} = \frac{1}{2\pi} \sum_{n=-\infty}^{+\infty} \sum_{m=-\infty}^{+\infty} H_{m-\ell}^{(1)}(kd) \int_0^{2\pi} J_m(kA_{\theta,1}) e^{i(m-n)\theta} d\theta, \quad (22)$$

$$\psi_{\ell'n'}^{TM} = \frac{1}{2\pi} \sum_{n'=-\infty}^{+\infty} i^{-n'} e^{-in'\alpha} e^{-ikd\cos\alpha} \int_0^{2\pi} J_{n'}(kA_{\theta,2}) e^{i(n'-\ell')\theta'} d\theta', \quad (23)$$

$$\Omega_{\ell'n'}^{TM} = \frac{1}{2\pi} \sum_{n'=-\infty}^{+\infty} \int_0^{2\pi} H_{n'}^{(1)}(kA_{\theta,2}) e^{i(n'-\ell')\theta'} d\theta', \quad (24)$$

$$\Omega_{\ell'n'm'}^{TM} = \frac{1}{2\pi} \sum_{n'=-\infty}^{+\infty} \sum_{m'=-\infty}^{+\infty} H_{\ell'-m'}^{(1)}(kd) \int_0^{2\pi} J_{m'}(kA_{\theta,2}) e^{i(m'-n')\theta'} d\theta'. \quad (25)$$

The solution of the coupled systems of equations given by Eqs.(9) and (19) for the TE and TM polarization cases, respectively, can be accomplished numerically by developing a program code in order to determine the scattering coefficients of the two particles. It must be emphasized that the numerical integration in Eqs.(10)-(15) and (20)-(25) should be performed prior to the summations of the partial-wave series.

### 3. Longitudinal and transverse radiation force functions

As demonstrated in [1] using the partial-wave series expansion method in cylindrical coordinates for the circular cylinder pair case, each object experiences an EM radiation force vector with two components; one longitudinal along the *x*- or *x'*-axis, and one transversal along the *y*- or *y'*-axis. The radiation force components are expressed in terms of longitudinal and transverse dimensionless functions $Y_{x,\{1,2\}}$ and $Y_{y,\{1,2\}}$ [1], which correspond to the radiation force per unit characteristic energy density factor ($I_0/c$) and geometrical cross-sectional area of the particle. Following the previous analysis [1], longitudinal and transverse dimensionless radiation force functions are derived for the pair of particles of arbitrary shapes. Their series expressions are given as,

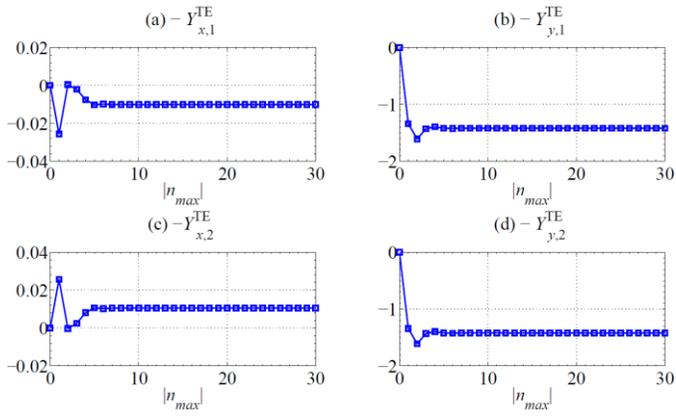

**Fig. 7.** The same as in Fig. 3, but for a pair of elliptical cylinders with smooth surfaces as shown in panel (e) of Fig. 2.

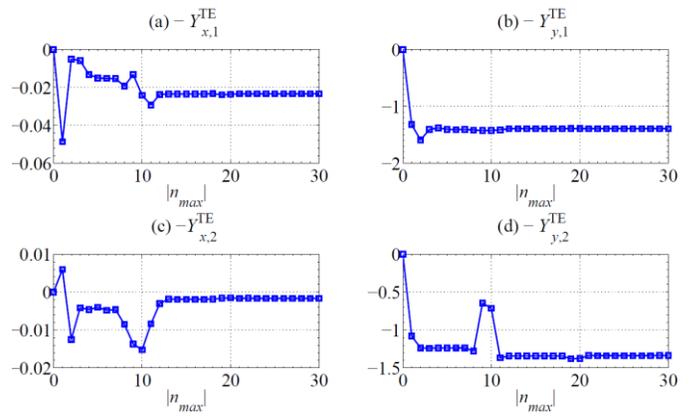

**Fig. 10.** The same as in Fig. 3, but for a smooth elliptical and corrugated cylinder pair as shown in panel (h) of Fig. 2.

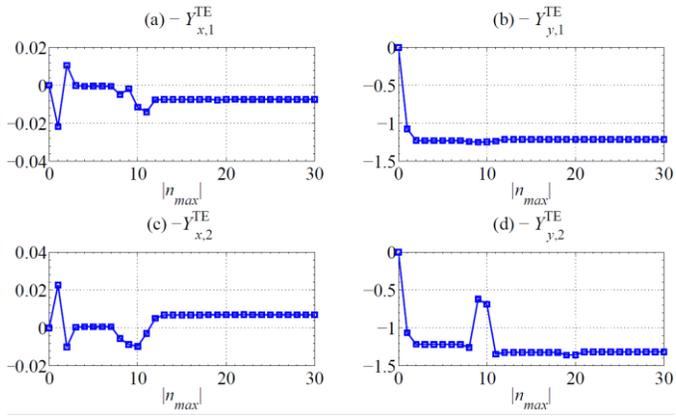

**Fig. 8.** The same as in Fig. 3, but for a smooth circular and corrugated cylinder pair as shown in panel (f) of Fig. 2.

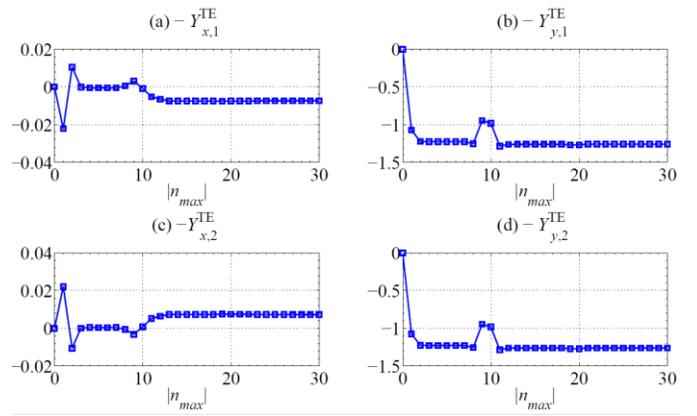

**Fig. 11.** The same as in Fig. 3, but for a pair of corrugated cylinder pair as shown in panel (i) of Fig. 2.

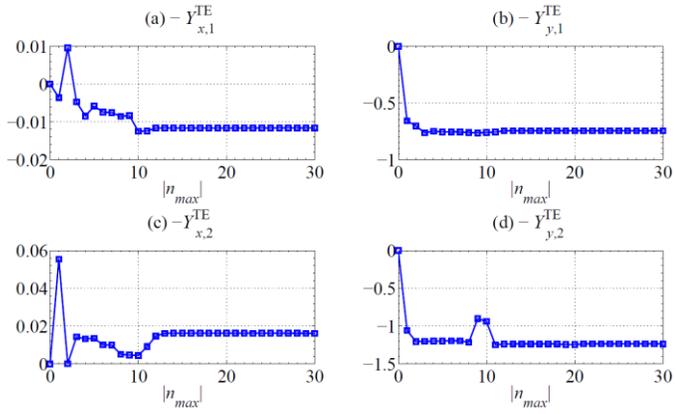

**Fig. 9.** The same as in Fig. 3, but for an elliptical with a smooth surface and a corrugated cylinder pair as shown in panel (g) of Fig. 2.

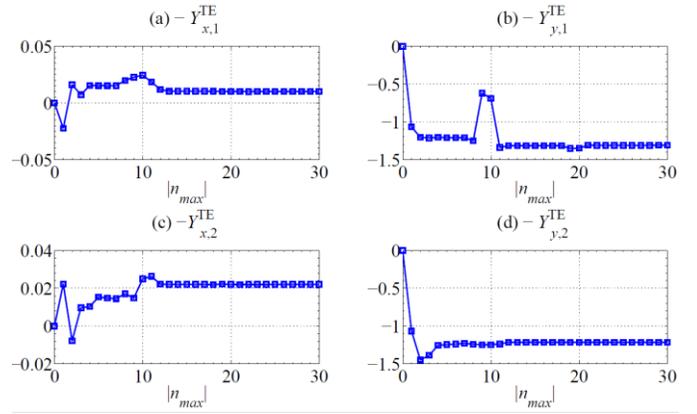

**Fig. 12.** The same as in Fig. 3, but for a corrugated cylinder and another cylinder of arbitrary shape as shown in panel (j) of Fig. 2.

$$Y_{x,1}^{\{TE,TM\}} = \frac{1}{ka_1} \Im\left\{\sum_{n=-\infty}^{+\infty} C_{n,1}^{\{TE,TM\}} \left[ \begin{array}{l} C_{n+1,1}^{\{TE,TM\}*} + i^{n+1}e^{i(n+1)\alpha} + \left(\sum_{m=-\infty}^{+\infty} C_{m,2}^{\{TE,TM\}} H_{n-m+1}(kd)\right)^* \\ -C_{n-1,1}^{\{TE,TM\}*} - i^{n-1}e^{i(n-1)\alpha} - \left(\sum_{m=-\infty}^{+\infty} C_{m,2}^{\{TE,TM\}} H_{n-m-1}(kd)\right)^* \end{array} \right] \right\}, \quad (26)$$

$$Y_{y,1}^{\{TE,TM\}} = -\frac{1}{ka_1} \Re\left\{\sum_{n=-\infty}^{+\infty} C_{n,1}^{\{TE,TM\}} \left[ \begin{array}{l} C_{n+1,1}^{\{TE,TM\}*} + i^{n+1}e^{i(n+1)\alpha} + \left(\sum_{m=-\infty}^{+\infty} C_{m,2}^{\{TE,TM\}} H_{n-m+1}(kd)\right)^* \\ +C_{n-1,1}^{\{TE,TM\}*} + i^{n-1}e^{i(n-1)\alpha} + \left(\sum_{m=-\infty}^{+\infty} C_{m,2}^{\{TE,TM\}} H_{n-m-1}(kd)\right)^* \end{array} \right] \right\}, \quad (27)$$

$$Y_{x,2}^{\{TE,TM\}} = \frac{1}{ka_2} \Im\left\{\sum_{n=-\infty}^{+\infty} C_{n,2}^{\{TE,TM\}} \left[ \begin{array}{l} C_{n+1,2}^{\{TE,TM\}*} + i^{n+1}e^{i(n+1)\alpha}e^{ikd\cos\alpha} + \left(\sum_{m=-\infty}^{+\infty} C_{m,1}^{\{TE,TM\}} H_{m-n-1}(kd)\right)^* \\ -C_{n-1,2}^{\{TE,TM\}*} - i^{n-1}e^{i(n-1)\alpha}e^{ikd\cos\alpha} - \left(\sum_{m=-\infty}^{+\infty} C_{m,1}^{\{TE,TM\}} H_{m-n+1}(kd)\right)^* \end{array} \right] \right\}, \quad (28)$$

$$Y_{y,2}^{\{TE,TM\}} = -\frac{1}{ka_2} \Re\left\{\sum_{n=-\infty}^{+\infty} C_{n,2}^{\{TE,TM\}} \left[ \begin{array}{l} C_{n+1,2}^{\{TE,TM\}*} + i^{n+1}e^{i(n+1)\alpha}e^{ikd\cos\alpha} + \left(\sum_{m=-\infty}^{+\infty} C_{m,1}^{\{TE,TM\}} H_{m-n-1}(kd)\right)^* \\ +C_{n-1,2}^{\{TE,TM\}*} + i^{n-1}e^{i(n-1)\alpha}e^{ikd\cos\alpha} + \left(\sum_{m=-\infty}^{+\infty} C_{m,1}^{\{TE,TM\}} H_{m-n+1}(kd)\right)^* \end{array} \right] \right\}, \quad (29)$$

where the symbols $\Re\{\cdot\}$ and $\Im\{\cdot\}$ denote the real and imaginary part of a complex number, respectively.

## 4. Computational results and discussions

The computational analysis is concentrated on evaluating the dimensionless radiation force functions for electrically-conducting cylinders in TE and TM polarized plane waves with arbitrary incidence angle $\alpha$ as given by Eqs.(26)-(29), for the ten examples of particle pairs shown in panels (a)-(j) in Fig. 2, respectively. Particles shapes deviating from the circular form are considered, as well as smooth surfaces [panels (a)-(e)] including corrugations [panels (f)-(j)].

The developed numerical code allows computing the scattering coefficients used subsequently to compute the radiation force components. Those coefficients are determined using matrix inversion procedures (similar to those used previously in the acoustical context [7, 9, 10]). Moreover, the related integrals are evaluated using the trapezoidal rule for numerical integration with a sampling of 650 points of the surface that ensures convergence.

Initially, convergence curves are evaluated in order to determine the maximum truncation limit in the series $|n_{max}|$ that provide negligible truncation error. Such plots are necessary to provide the required accuracy and ensure the correctness of the results.

Panels (a)-(d) in Figs. 3-12 display the convergence plots for the longitudinal and transverse radiation force functions for the particle pair assuming TE polarization of the incident plane waves with $\alpha = 90°$ and $kd = 30$. The maximum truncation limit has been varied in range $0 \leq |n_{max}| \leq 30$. Depending on each particle pair considered according to Fig. 2, suitable convergence is achieved differently.

Panels (a)-(d) of Fig. 3 correspond to the particle pair shown in panel (a) of Fig. 2, such that $(ka_1/kb_1) = (ka_2/kb_2) = 1$, and $d_1 = d_2 = \ell_1 = \ell_2 = 0$. All the plots for the longitudinal and transverse radiation force functions evaluated for $\alpha = 90°$ and $kd = 30$ show that there is no contribution of the monopole mode $n = 0$, and the convergence to the stable/steady solution is achieved for $|n_{max}| = 4$.

As the shape of particle 2 deviates from the perfect circular cylinder such that $(ka_1/kb_1) = 1$, $(ka_2/kb_2) = 1.5$ while $d_1 = d_2 = \ell_1 = \ell_2 = 0$, convergence of the longitudinal and transverse radiation force functions has been attained at a larger truncation limit as shown in Fig. 4. Panel (a) for $Y_{x,1}^{TE}$ shows that the convergence is attained for $|n_{max}| = 7$, while panel (b) for $Y_{y,1}^{TE}$ shows that a smaller truncation order is needed; i.e. $|n_{max}| = 4$. This is not the case for the plots shown in panels (c) and (d) where $|n_{max}| = 11$ is needed for $Y_{x,2}^{TE}$ while $|n_{max}| = 7$ is required for $Y_{y,2}^{TE}$. These plots demonstrate that each radiation force function converges differently.

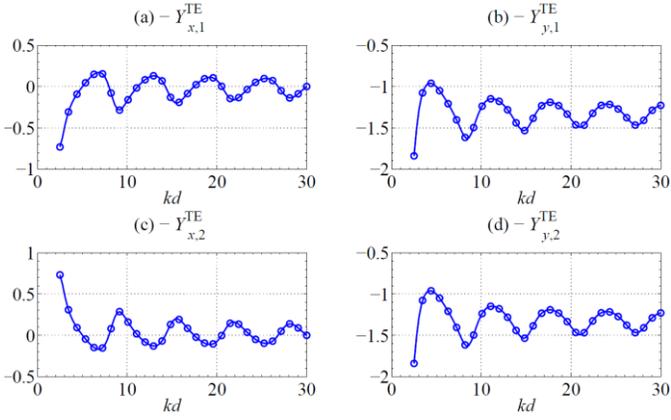

**Fig. 13.** Panels (a)–(d) display the plots for the longitudinal and transverse radiation force functions for two perfectly conducting circular cylinders as shown in panel (a) of Fig. 2 versus the dimensionless interparticle distance $kd$ for an angle of incidence $\alpha = 90°$.

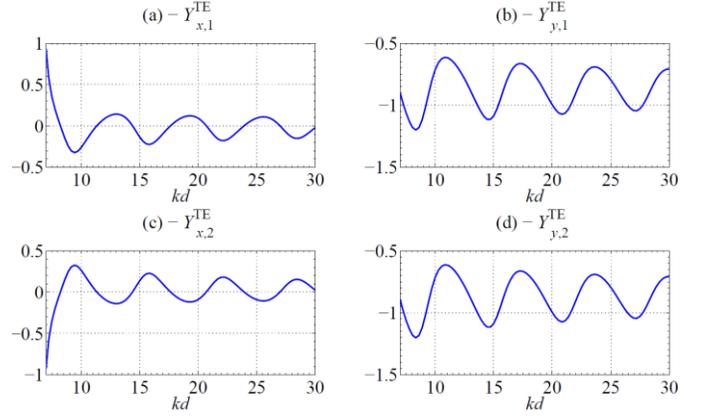

**Fig. 16.** The same as in Fig. 13, but the plots correspond to the radiation force functions for the cylinder pair shown in panel (d) of Fig. 2.

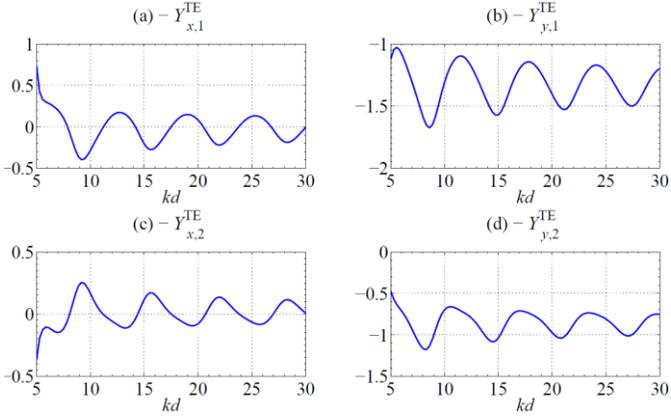

**Fig. 14.** The same as in Fig. 13, but the plots correspond to the radiation force functions for the cylinder pair shown in panel (b) of Fig. 2.

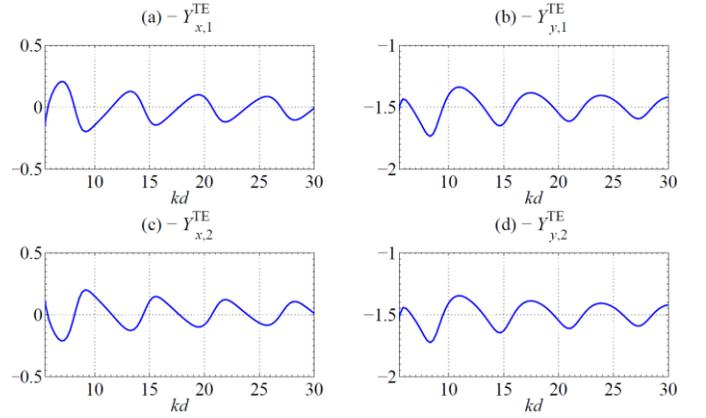

**Fig. 17.** The same as in Fig. 13, but the plots correspond to the radiation force functions for the cylinder pair shown in panel (e) of Fig. 2.

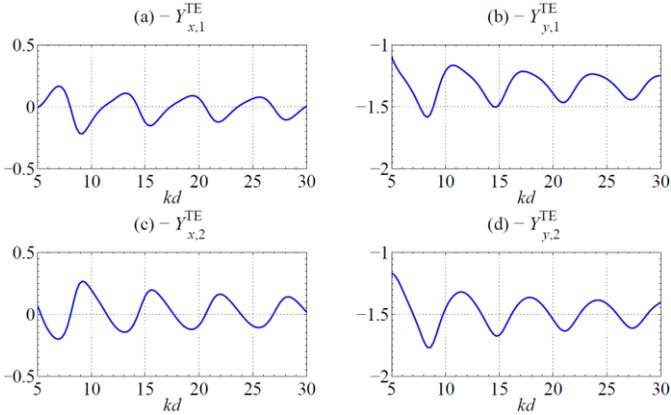

**Fig. 15.** The same as in Fig. 13, but the plots correspond to the radiation force functions for the cylinder pair shown in panel (c) of Fig. 2.

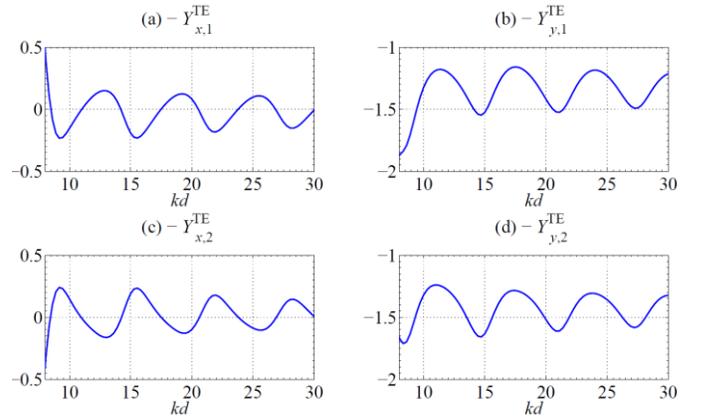

**Fig. 18.** The same as in Fig. 13, but the plots correspond to the radiation force functions for the cylinder pair shown in panel (f) of Fig. 2.

Consider now another particle pair as shown in panel (c) of

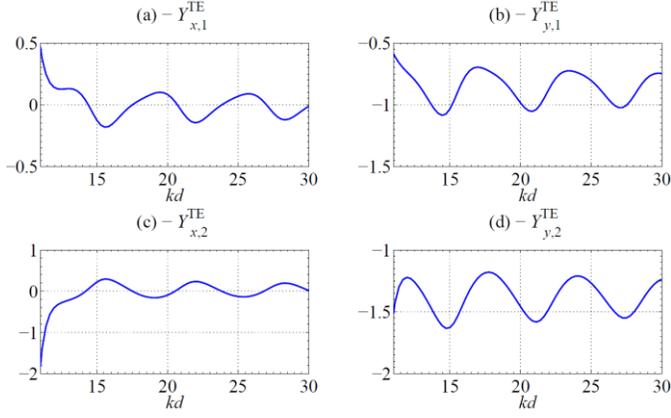

**Fig. 19.** The same as in Fig. 13, but the plots correspond to the radiation force functions for the cylinder pair shown in panel (g) of Fig. 2.

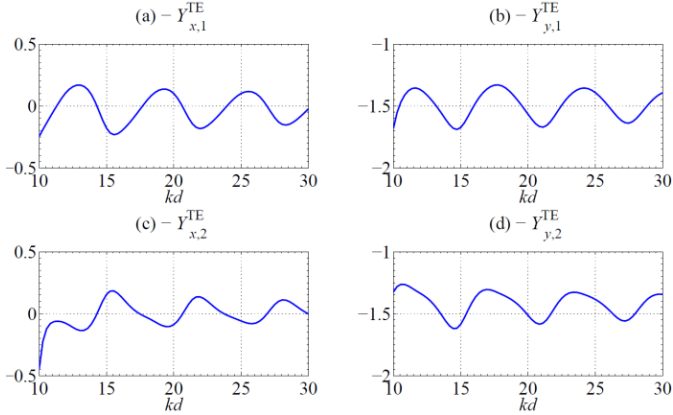

**Fig. 20.** The same as in Fig. 13, but the plots correspond to the radiation force functions for the cylinder pair shown in panel (h) of Fig. 2.

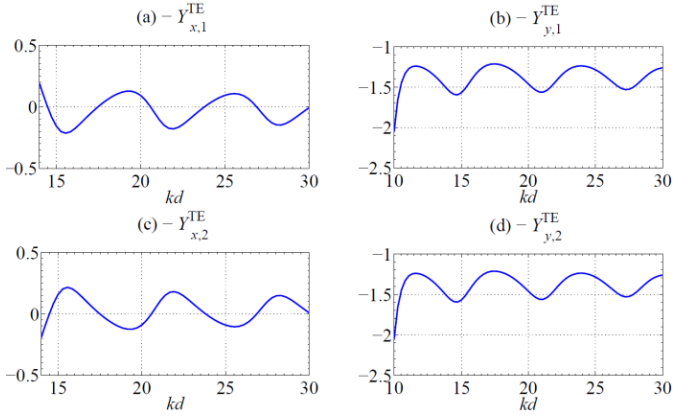

**Fig. 21.** The same as in Fig. 13, but the plots correspond to the radiation force functions for the cylinder pair shown in panel (i) of Fig. 2.

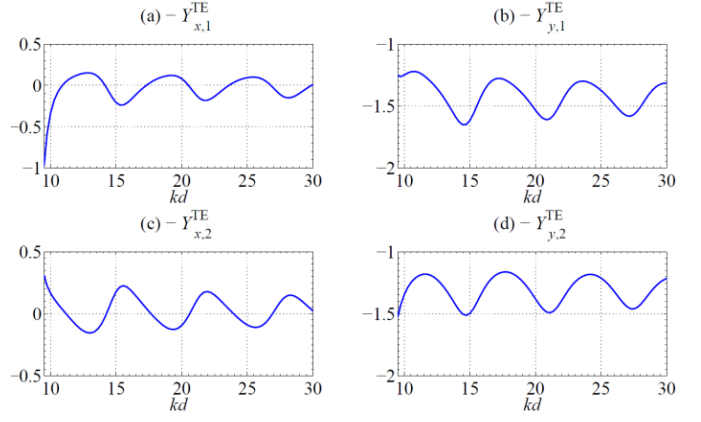

**Fig. 22.** The same as in Fig. 13, but the plots correspond to the radiation force functions for the cylinder pair shown in panel (j) of Fig. 2.

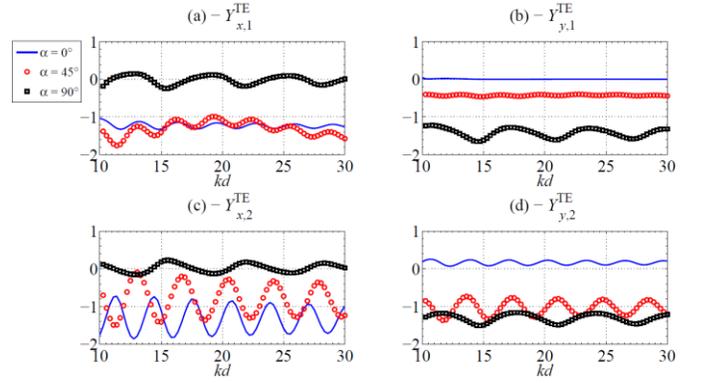

**Fig. 23.** Panels (a)–(d) display the plots for the longitudinal and transverse radiation force functions for the cylinder pair shown in panel (j) of Fig. 2 versus the dimensionless interparticle distance $kd$ for different incidence angles assuming TE polarization of the incident plane wave field.

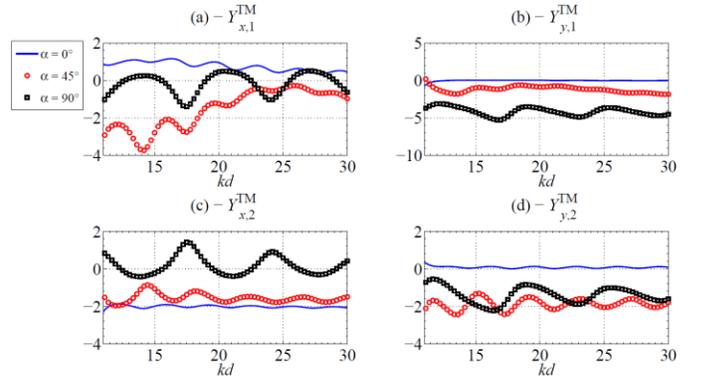

**Fig. 24.** The same as in Fig. 23, but a TM polarized field is considered.

Fig. 2, such that $(ka_1/kb_1) = 1$, $(ka_2/kb_2) = 0.75$ while $d_1 = d_2 = \ell_1 = \ell_2 = 0$. Fig. 5 displays the corresponding plots where convergence is obtained for a maximum limit $|n_{max}| = 7$ for all the curves.

The case of a pair of elliptical cylinders with smooth surfaces is now considered, such that $(ka_1/kb_1) = (ka_2/kb_2) = 1.5$ while $d_1 = d_2 = \ell_1 = \ell_2 = 0$, corresponding to the geometry described in panel (d) of Fig. 2. Fig. 6 displays the related convergence plots where the steady limit is obtained for a maximum truncation $|n_{max}| = 12$ for all the curves.

Another form of elliptical cylinders has been also considered such that $(ka_1/kb_1) = (ka_2/kb_2) = 0.75$ while $d_1 =$

$d_2 = \ell_1 = \ell_2 = 0$, corresponding to the geometry described in panel (e) of Fig. 2. In contrast with the plots shown in the panels of Fig. 6, the convergence in the curves for the elliptical cylinder pair in Fig. 7 corresponding to the configuration shown in panel (e) of Fig. 2 requires less partial waves such that the maximum limit $|n_{max}| = 6$.

Consider now a pair of cylinders, such that particle 1 has a circular shape while particle 2 has a corrugated/ribbed surface such that $(ka_1/kb_1) = (ka_2/kb_2) = 1$, while $d_1 = \ell_1 = 0$, $d_2 = 0.1\,a_2$, and $\ell_2 = 10$, as shown in panel (f) of Fig. 2. Panels (a)-(d) of Fig. 8 display the convergence plots; panels (a) and (b) for $Y_{x,1}^{TE}$ and $Y_{y,1}^{TE}$, respectively, show that the convergence is obtained for the circular cylinder at a truncation limit $|n_{max}| = 12$, while panel (c) for $Y_{x,2}^{TE}$ shows that a larger truncation order is needed; i.e. $|n_{max}| = 14$. This is not the case for the plot shown in panel (d) where $|n_{max}| = 21$ is needed for $Y_{y,2}^{TE}$.

In the next example, an elliptical cylinder with a smooth surface in addition to a corrugated cylinder are considered such that $(ka_1/kb_1) = 1.5$, $(ka_2/kb_2) = 1$, while $d_1 = \ell_1 = 0$, $d_2 = 0.05\,a_2$, and $\ell_2 = 10$, as displayed in panel (g) of Fig. 2. The corresponding convergence plots for the longitudinal and transverse radiation force functions are displayed in panels (a)-(d) of Fig. 9. Similar to Fig. 8, panels (a) and (b) of Fig. 9 for $Y_{x,1}^{TE}$ and $Y_{y,1}^{TE}$, respectively, show that the convergence is obtained at a truncation limit $|n_{max}| = 12$ for both curves. Panel (c) for $Y_{x,2}^{TE}$ shows that a larger truncation order is needed; i.e. $|n_{max}| = 13$, while the plot shown in panel (d) suggests a larger maximum limit is needed for which $|n_{max}| = 21$ is needed for $Y_{y,2}^{TE}$.

Next, a different geometry for the particle pair is considered such that an elliptical cylinder with a smooth surface in addition to a corrugated cylinder are considered with the following chosen parameters; $(ka_1/kb_1) = 0.75$, $(ka_2/kb_2) = 1$, while $d_1 = \ell_1 = 0$, $d_2 = 0.1\,a_2$, and $\ell_2 = 10$, as displayed in panel (h) of Fig. 2. For this example, the convergence plots are displayed in panels (a)-(d) of Fig. 10, where a maximum truncation limit $|n_{max}| = 21$ is required.

The case of two similar corrugated cylinders is now considered such that $(ka_1/kb_1) = (ka_2/kb_2) = 1$, while $(d_1, d_2) = 0.05(a_1, a_2)$ and $\ell_1 = \ell_2 = 10$, as shown in panel (i) of Fig. 2. Panels (a)-(d) of Fig. 11 display the corresponding convergence plots where panels (a) and (c) for $Y_{x,1}^{TE}$ and $Y_{x,2}^{TE}$, respectively, show that the convergence is obtained for $|n_{max}| = 14$. Panels (b) and (d) for $Y_{y,1}^{TE}$ and $Y_{y,2}^{TE}$ show that a larger truncation limit is needed such that $|n_{max}| = 21$.

Finally, the example of a corrugated cylinder with an irregularly shaped one is considered such that $(ka_1/kb_1) = (ka_2/kb_2) = 1$, while $(d_1, d_2) = 0.1(a_1, a_2)$, $\ell_1 = 10$ and $\ell_2 = 3$, as shown in panel (j) of Fig. 2. Panels (a), (c) and (d) of Fig. 12 for $Y_{x,1}^{TE}$, $Y_{x,2}^{TE}$ and $Y_{y,2}^{TE}$, respectively, show that the convergence is obtained for $|n_{max}| = 13$, while panel (b) $Y_{y,1}^{TE}$ shows that a larger truncation limit is needed to ensure convergence, such that $|n_{max}| = 23$.

Next, computations for $Y_{\{x,y\},\{1,2\}}^{\{TM,TE\}}$ are performed in the range $0 < kd \le 30$ with emphasis on varying the shapes of the cylindrical particle pairs shown in panels (a)-(j) of Fig. 2. Based on the convergence curves computed previously, the maximum truncation limit which ensures the required accuracy of the results is set to $|n_{max}| = 25$ and $\alpha = 90°$.

Panels (a)-(d) of Fig. 13 display the results for the radiation force functions $Y_{\{x,y\},\{1,2\}}^{TE}$ for the electrically-conducting circular cylinder pair [shown in panel (a) of Fig. 2], such that $(ka_1/kb_1) = (ka_2/kb_2) = 1$, and $d_1 = d_2 = \ell_1 = \ell_2 = 0$. The circle markers in all the panels correspond to the computations performed using the previously developed theoretical analysis for circular cylinders [1], where complete agreement has been obtained. This test provides some confidence (in addition to the convergence plots computed and discussed previously) on the accuracy and validity of the results. In panels (a) and (c), the longitudinal radiation force functions oscillate between positive and negative values, suggesting electromagnetic binding (i.e., attraction between particles) or radiation force reversal (i.e., repulsion between the particles) depending on the value of $kd$. Furthermore, the transverse radiation force functions in panels (b) and (d) are negative. There are values of $kd$ where $Y_{x,\{1,2\}}^{TE}$ vanishes such that the corresponding cylinder experiences no force, and becomes irresponsive to the transfer of linear momentum.

Panels (a)-(d) of Fig. 14 correspond to the electrically-conducting cylinder pair as shown in panel (b) of Fig. 2; i.e. a circular with an elliptical cylinder such that $(ka_1/kb_1) = 1$, $(ka_2/kb_2) = 1.5$ while $d_1 = d_2 = \ell_1 = \ell_2 = 0$. Deviation from the circular shape of one of the cylinders alters the behavior of the force functions, nonetheless, the variations of the longitudinal components $Y_{x,\{1,2\}}^{TE}$ between positive and negative values remain manifested in the plots, suggesting EM binding and force reversal as $kd$ varies.

Figs. 15-22 display the computational results for the radiation force functions $Y_{\{x,y\},\{1,2\}}^{TE}$ for the particle pairs shown in panels (c)-(j) of Fig. 2 using the same parameters $(ka_1/kb_1)$, $(ka_2/kb_2)$, $d_1$, $d_2$, $\ell_1$ and $\ell_2$ used to compute the convergence plots. As a general observation, the longitudinal radiation force functions are large as $kd$ decreases, especially for the corrugated cylinder pair shown in Fig. 21. Oscillations between positive and negative values still occur regardless of particle shape, as well as a vanishing of $Y_{x,\{1,2\}}^{\text{TE}}$ at particular values of *kd*.

The effect of changing the incidence angle is also investigated for the particle pair shown in panel (j) of Fig. 2, for three different values of $\alpha = 0°$, 45° and 90°. The corresponding results are displayed in panels (a)-(d) of Fig. 23, where the behavior of the longitudinal and transverse components is altered significantly. The oscillations between positive and negative values versus *kd* observed for $\alpha = 90°$, are now altered such that the plots for $Y_{x,\{1,2\}}^{\text{TE}}$ are always negative for $\alpha = 0°$ and $\alpha = 45°$.

Changing the polarization of the incident electric field to TM also affects the behavior of the radiation force functions, as shown in panels (a)-(d) of Fig. 24 for the particle pair shown in panel (j) of Fig. 2. Comparison of panel (a) of Fig. 24 with that of Fig. 23 shows that for $\alpha = 0°$, the longitudinal radiation force function changes sign. Thus, these results suggest that it is possible to alter the optical binding or repulsion between a pair of particles by switching the polarization of the incident field from TE to TM or vice-versa.

## 5. Conclusion and perspectives

In summary, the present investigation provided a semi-analytical approach to compute the EM radiation force-per-length exerted on a pair of electrically-conducting cylindrical particles of arbitrary geometrical cross-sections based on boundary matching in cylindrical coordinates. Adequate convergence plots confirm the validity of the method to evaluate the radiation force functions. Moreover, the radiation force functions oscillate between positive and negative values depending on the interparticle distance separating their centers of mass, the angle of incidence as well as the polarization of the electric field. It is noteworthy that the analytical theory and formalism developed here is adaptable to any frequency range (i.e. Rayleigh, Mie or geometrical optics regimes). The results can find potential applications in optical tweezers and other related applications in fluid dynamics to study the deformation [31, 32] and stability of liquid bridges in electric fields [33, 34], opto-fluidic devices, and particle manipulation to name a few examples.

Concerning the acoustical analog, it is important to note that the exact closed-form expressions and corresponding results presented here are directly applicable to the case of rigid (TE) or soft (TM) cylindrical particle pairs with smooth or corrugated surfaces. The radiation force functions given by Eqs.(26)-(29) are equivalent to those developed and presented previously[3]. Therefore, one can conclude the acoustic radiation force behavior from the results presented here assuming rigid or soft particle pairs of arbitrary shapes, as shown in panels (a)-(j) of Fig. 2.

[27] Gong Z, Li W, Mitri FG, Chai Y, Zhao Y. Arbitrary scattering of an acoustical Bessel beam by a rigid spheroid with large aspect-ratio. *J Sound Vib* 2016;383:233-47.
[28] Gong Z, Li W, Chai Y, Zhao Y, Mitri FG. T-matrix method for acoustical Bessel beam scattering from a rigid finite cylinder with spheroidal endcaps. *Ocean Engineering* 2017;129:507-19.
[29] Mitri FG. Axial acoustic radiation force on rigid oblate and prolate spheroids in Bessel vortex beams of progressive, standing and quasi-standing waves. *Ultrasonics* 2017;74:62-71.
[30] Acoustic, electromagnetic and elastic wave scattering - Focus on the *T*-matrix approach. (V. K. Varadan and V. V. Varadan, Eds.) NY: Pergamon; 1980.
[31] Dodgson N, Sozou C. The deformation of a liquid drop by an electric field. *Z angew Math Phys* 1987;38:424-32.
[32] Deshmukh SD, Thaokar RM. Deformation, breakup and motion of a perfect dielectric drop in a quadrupole electric field. *Phys Fluids* 2012;24:032105.
[33] Sankaran S, Saville DA. Experiments on the stability of a liquid bridge in an axial electric field. *Physics of Fluids A: Fluid Dynamics* 1993;5:1081-3.
[34] Mhatre S. Dielectrophoretic motion and deformation of a liquid drop in an axisymmetric non-uniform AC electric field. *Sensors and Actuators B: Chemical* 2017;239:1098-108.
12